\documentclass{ws-procs9x6}            

\newcommand*\diff{\mathop{}\!\mathrm{d}}

\newcommand{\nn}{\nonumber}

\newcommand{\be}{\begin{eqnarray}}
\newcommand{\ee}{\end{eqnarray}}

\newcommand{\ml}{\mathcal}
\newcommand{\bs}{\boldsymbol}

\begin{document}
\title{Fate of Heavy Quark Bound States inside Quark-Gluon Plasma} 

\author{Xiaojun Yao$^*$, Weiyao Ke, Yingru Xu, Steffen Bass, Thomas Mehen and Berndt M\"uller}

\address{Department of Physics, Duke University, Durham, NC 27708, USA\\
$^*$Center for Theoretical Physics, Massachusetts Institute of Technology\\
Cambridge, MA 02139, USA\\
$^*$E-mail: xiaojun.yao@duke.edu}

\begin{abstract}
Transport equations have been applied successfully to describe the quarkonium evolution inside the quark-gluon plasma, which include both plasma screening effects and recombination. We demonstrate how the quarkonium transport equation is derived from QCD by using the open quantum system framework and effective field theory. Weak coupling and Markovian approximations used in the derivation are justified from a separation of scales. By solving the equations numerically, we study the Upsilon production in heavy ion collisions.
\end{abstract}


\preprint{MIT-CTP-5161}

\bodymatter

\section{Introduction}\label{aba:sec1}

Heavy quarkonium is a bound state of a heavy quark-antiquark pair ($Q\bar{Q}$). Inside the quark-gluon plasma (QGP), which is a deconfined phase of nuclear matter, the attractive potential between the $Q\bar{Q}$ is screened. The plasma screening becomes stronger as the temperature increases. As a result, at sufficiently high temperature, the potential is too weak to support the bound state formation \cite{Matsui:1986dk,Karsch:1987pv}. In other words, quarkonium ``melts". This effect is called static plasma screening effect. It leads to suppression of quarkonium production in heavy ion collisions, if the QGP is formed. Thus, quarkonium suppression is used as a signature of the QGP formation in heavy ion collisions. 

Experimental measurements confirmed the quarkonium suppression in heavy ion collisions at both RHIC and LHC. To explain the data quantitatively, one has to take into account another plasma screening effect, called dynamical screening effect. It describes the quarkonium dissociation in dynamical scattering processes. It happens when enough energy is transferred from the medium to the quarkonium state to excite it to the continuum. It generates a thermal width of quarkonium, which increases with the temperature. These two screening effects are connected in the sense that both of them are generated from the thermal loop correction to the quarkonium propagator.

In addition to the plasma screening effects, another crucial process in understanding the quarkonium in-medium evolution is the recombination of unbound $Q\bar{Q}$'s \cite{Thews:2000rj,Andronic:2007bi}. When the medium temperature is below the melting temperature of quarkonium, an unbound $Q\bar{Q}$ may radiate out a certain amount of energy and form a bound state. The $Q$ and $\bar{Q}$ in the recombination may come from the same initial hard scattering vertex or different initial hard vertices. The latter case is negligible when only a few heavy quarks are produced in the collision, but will be enhanced as the heavy quark number increases. The recombination is crucial to explain the collision energy dependence of J$/\psi$ suppression. J$/\psi$ is less suppressed as the collision energy increases from RHIC to LHC because of the enhanced recombination (see Ref.~\citenum{Andronic:2015wma} for a review).

To understand the quarkonium in-medium dynamics, one has to consider both plasma screening effects and recombination. Transport equations have been applied successfully \cite{Grandchamp:2003uw,Yan:2006ve,Krouppa:2015yoa,Du:2017qkv,Zhao:2017yan,Yao:2018zrg,Yao:2018dap}. However, the connection between the transport equation and the underlying theory QCD is not clear. Furthermore, it is not clear when the transport equation is a valid description. In this proceeding, we will explain how to derive the Boltzmann transport equation for quarkonium in-medium dynamics from QCD using the open quantum system framework. The transport equation will be briefly reviewed in Sect.~\ref{aba:sec2} and the derivation will be explained in Sect.~\ref{aba:sec3}. We will discuss the relation between the validity condition of transport equations and a hierarchy of scales. Finally, phenomenological results on Upsilon suppression will be shown in Sect.~\ref{aba:sec4} and conclusions will be drawn in Sect.~\ref{aba:sec5}.

\section{Boltzmann Transport Equation}\label{aba:sec2}
The Boltzmann transport equation for quarkonium can be written as:
\be
(\partial_t + \dot{{\bs x}}\cdot \nabla_{\bs x})f({\bs x}, {\bs k}, t) = \ml{C}^{+}({\bs x}, {\bs k}, t)  - \ml{C}^{-}({\bs x}, {\bs k}, t) \,,
\ee
in which $f({\bs x}, {\bs k}, t)$ is the phase space distribution function of a quarkonium state (each state has its own $f$). The left hand side describes the free streaming of quarkonium and $\dot{{\bs x}} \equiv \frac{\diff {\bs x}}{\diff t}$. The right hand side contains two collision terms: recombination $\ml{C}^{+}({\bs x}, {\bs k}, t)$ and dissociation $\ml{C}^{-}({\bs x}, {\bs k}, t)$. The static screening effect has been accounted for in the quarkonium wavefunction, which is used in the calculation of $\ml{C}^{\pm}$.

\section{Derivation of Transport Equation}\label{aba:sec3}
We will use the framework of open quantum system. In this framework, the whole system under study consists of a sub-system and an environment. The whole system evolves unitarily. The evolution is also time-reversible if the underlying theory respects time-reversal symmetry such as QCD. If we focus on just the sub-system, we will integrate out the environment degrees of freedom. Then the evolution equation of the sub-system is non-unitary and time-irreversible.
In our case, the sub-system consists of bound and unbound $Q\bar{Q}$'s while the environment is the QGP.

If we assume the sub-system and the environment interact weakly, we can expand the evolution equation to second order in perturbation and obtain the Lindblad equation:
\be \nn
\rho_S(t) &=& \rho_S(0) -i \Big[t H_S + \sum_{a,b} \sigma_{ab}(t) L_{ab}, \rho_S(0)  \Big]  \\
\label{eqn:lindblad}
&& + \sum_{a,b,c,d} \gamma_{ab,cd} (t) \Big( L_{ab}\rho_S(0)L^{\dagger}_{cd} - \frac{1}{2}\big\{ L^{\dagger}_{cd}L_{ab}, \rho_S(0) \big\}  \Big) \,,
\ee
where definitions of each term can be found in Ref.~\citenum{Yao:2018nmy}.
It can be shown that the Lindblad equation leads to the Boltzmann transport equation under the Markovian approximation if a Wigner transform is applied to the sub-system density matrix. The Wigner transform is defined by
\be
f({\bs x}, {\bs k}, t) \equiv \int\frac{\diff^3k'}{(2\pi)^3} e^{i {\bs k}'\cdot {\bs x} } \Big\langle  {\bs k}+\frac{{\bs k}'}{2}   \Big| \rho_S(t)  \Big|   {\bs k}-\frac{{\bs k}'}{2}  \Big\rangle\,,
\ee
which connects the density matrix and the phase space distribution. A one-to-one correspondence between terms in the Lindblad equation and those in the transport equation can be established. More specifically, the $\sigma_{ab}$ term gives the static screening; The $L_{ab}\rho_S(0)L^{\dagger}_{cd}$ term corresponds to recombination while the $\{ L^{\dagger}_{cd}L_{ab}, \rho_S(0)\}$ term leads to dissociation; Remaining terms give the free streaming. Detailed expressions of $\ml{C}^{\pm}$ can be found in Refs.~\citenum{Yao:2018nmy} and~\citenum{Yao:2018sgn}.

The weak coupling and Markovian approximations can be justified from a separation of scales: $M \gg Mv \gg Mv^2 \gtrsim T \gtrsim m_D$. Here $M$ denotes the heavy quark mass; $v$ is the relative velocity between the heavy quark pair inside quarkonium; $T$ is the plasma temperature and $m_D$ denotes the Debye mass. Under this hierarchy of scales, the effective field theory potential nonrelativistic QCD (pNRQCD) can be constructed from QCD by a sequence of renormalization group flow, matching calculations and nonrelativistic expansions \cite{Brambilla:1999xf,Fleming:2005pd}. In pNRQCD, a $Q\bar{Q}$ color singlet $S$ turns to an octet $O$ (and vice versa) by interacting with gluons in a dipole interaction: $O^\dagger {\bs r}\cdot {\bs E} S + h.c.$, where ${\bs E}$ is the chromo-electric field. Since the typical energy of medium gluons is $T$, the interaction vertex scales as $r T \sim \frac{T}{Mv} \sim v$ where $r$ is the quarkonium size. Since $v$ is assumed small, the interaction is weak. Furthermore, due to the weak interaction, the sub-system relaxation time can be shown to be much bigger than the environment correlation time, which justifies the Markovian approximation. The Markovian approximation is coarse graining, which means the finer detail of the environment dynamics is not resolved by the sub-system evolution.

In a nutshell, whenever the separation of scales $M \gg Mv \gg Mv^2 \gtrsim T \gtrsim m_D$ is true, the Boltzmann transport equation for quarkonium is valid. In practice, we have $Mv^2 \sim 500$ MeV for both charmonium and bottomonium and $T\lesssim 500$ MeV in current heavy ion experiments. The derivation here explains why the transport equation phenomenology works well. 

\section{Phenomenological Results}\label{aba:sec4}
In this section, we will show phenomenological results based on the quarkonium transport equation derived in the last section. After quarkonium dissociation, a valid description of the in-medium heavy quark dynamics is the transport equation of open heavy quarks \cite{Ke:2018tsh}. Thus, we have to couple the transport equations of quarkonium with those of open heavy quarks \cite{Yao:2017fuc}. We will focus on the bottomonium system since the separation of scales works better for the bottom quark.

We will solve the coupled transport equations by Monte Carlo simulations. Details can be found in Ref.~\citenum{Yao:2018zrg}.

\begin{figure}
\includegraphics[height=1.62in]{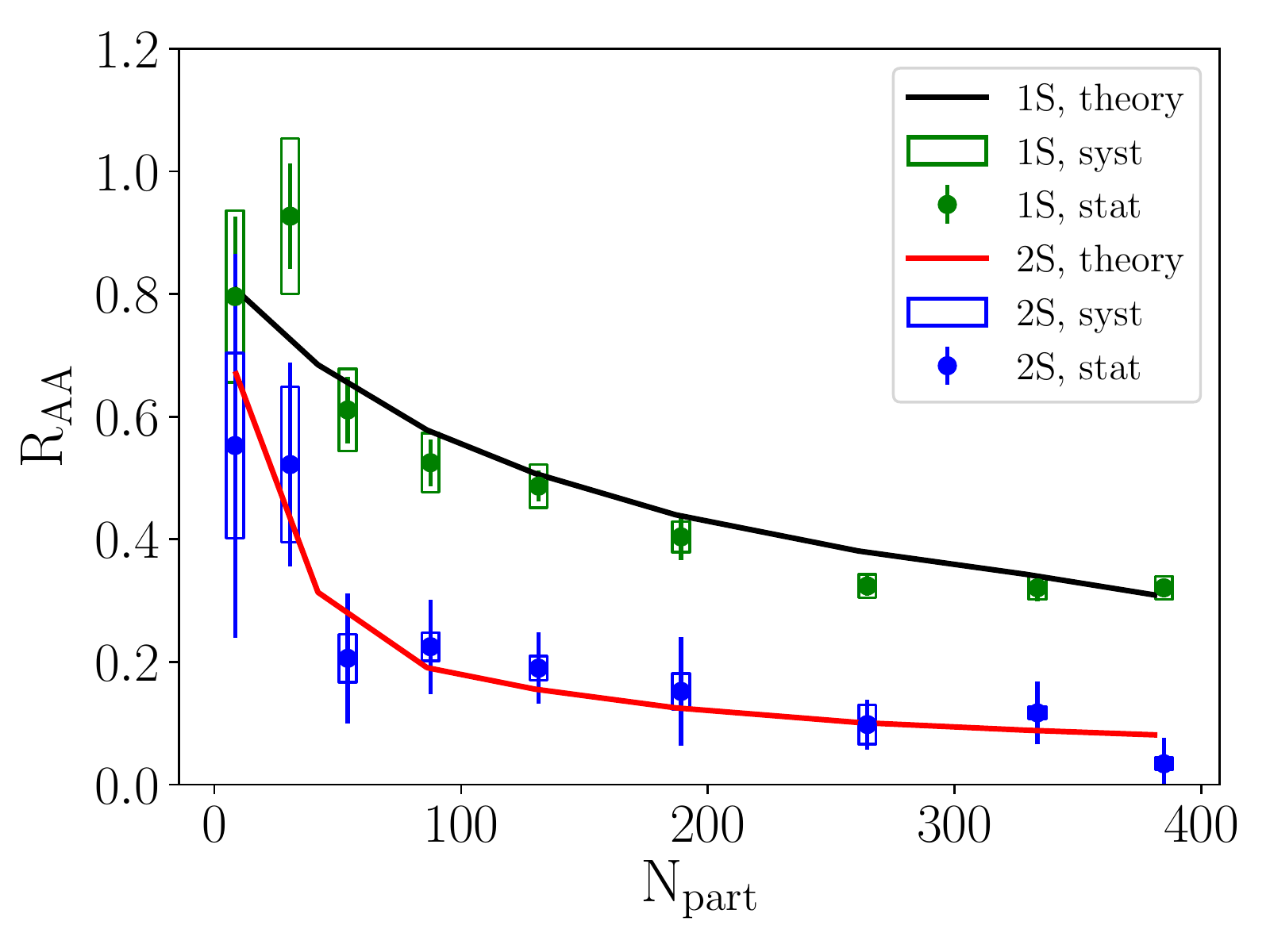}
\includegraphics[height=1.62in]{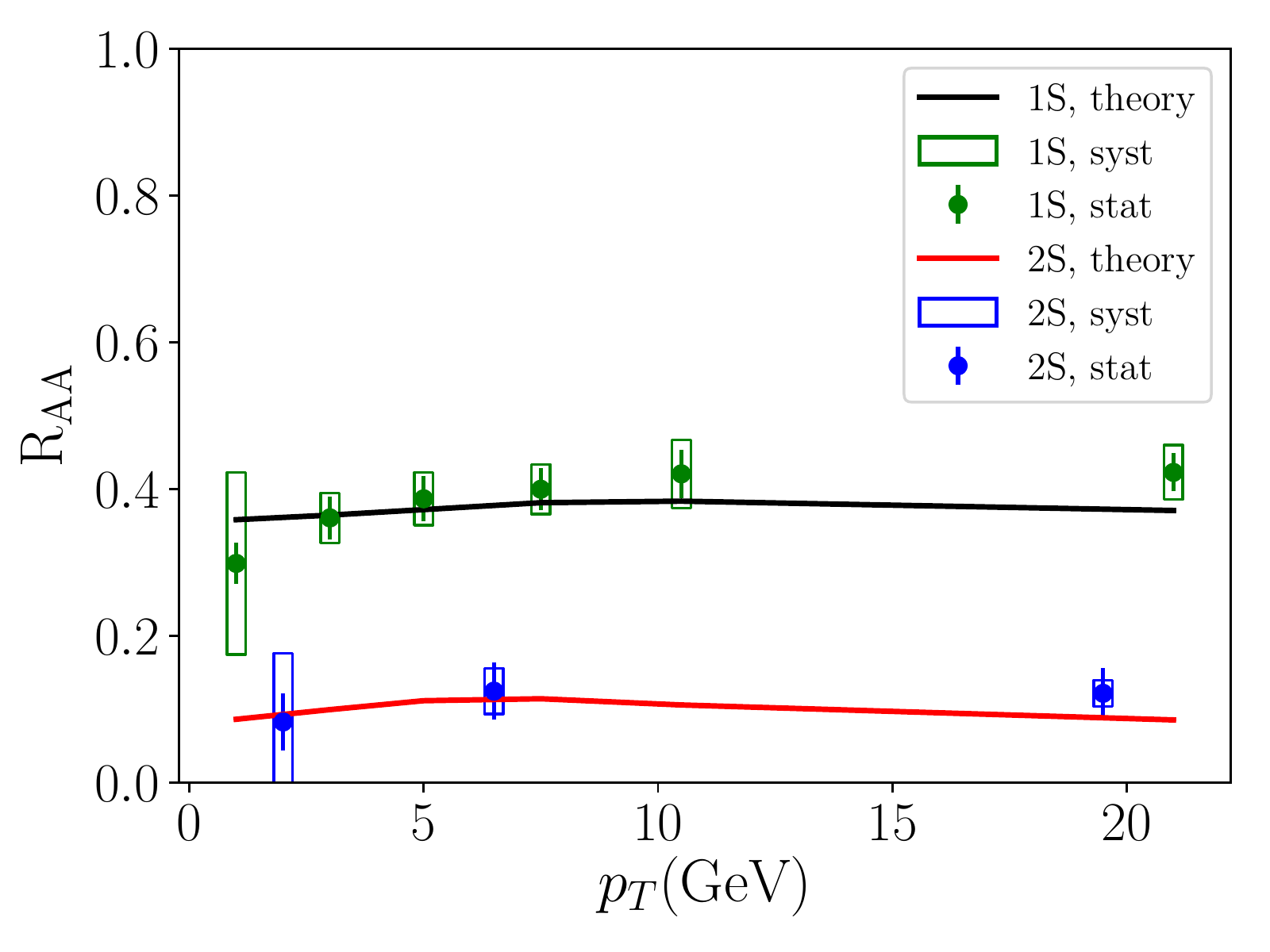}
\includegraphics[height=1.62in]{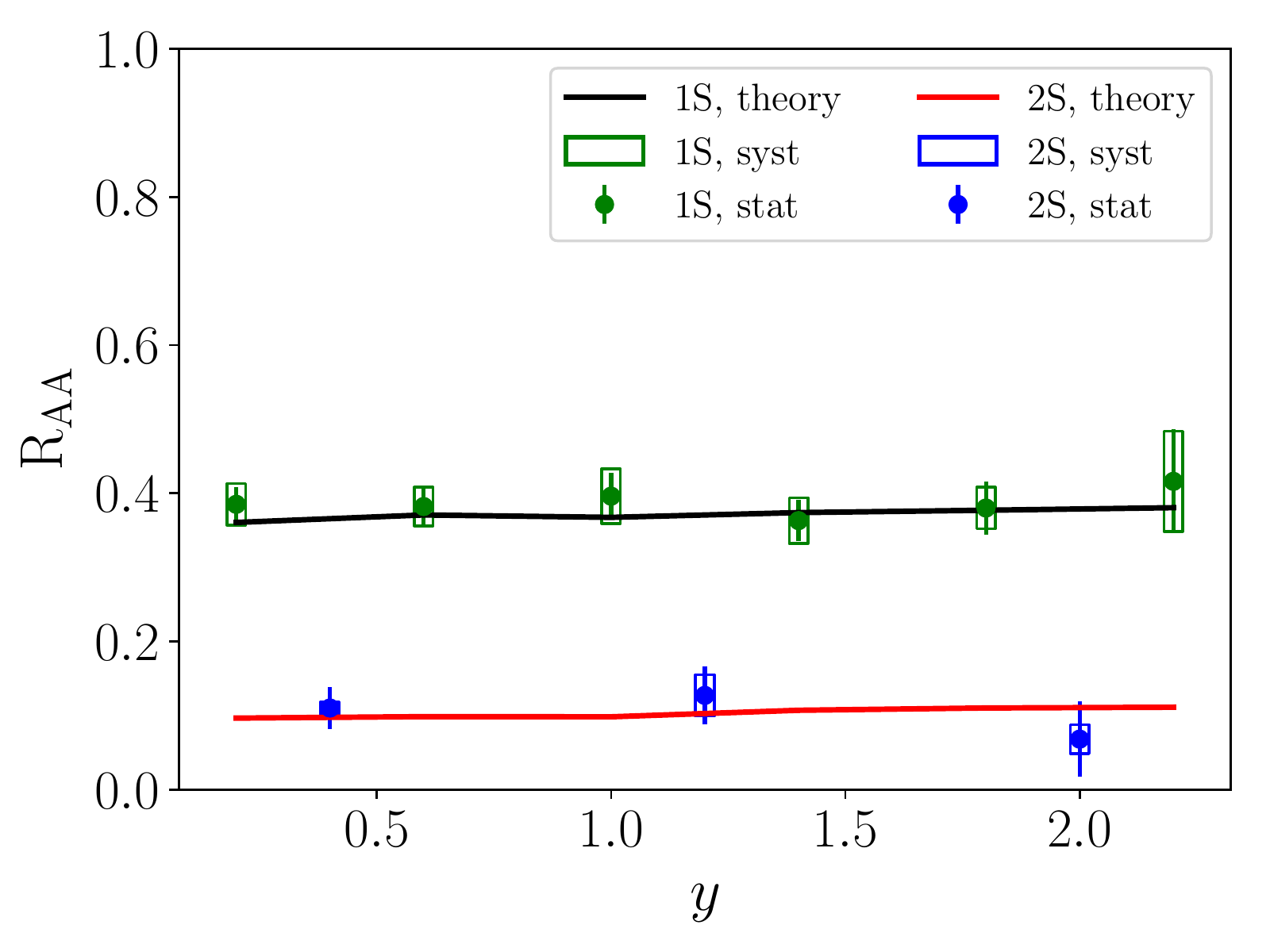}
\includegraphics[height=1.62in]{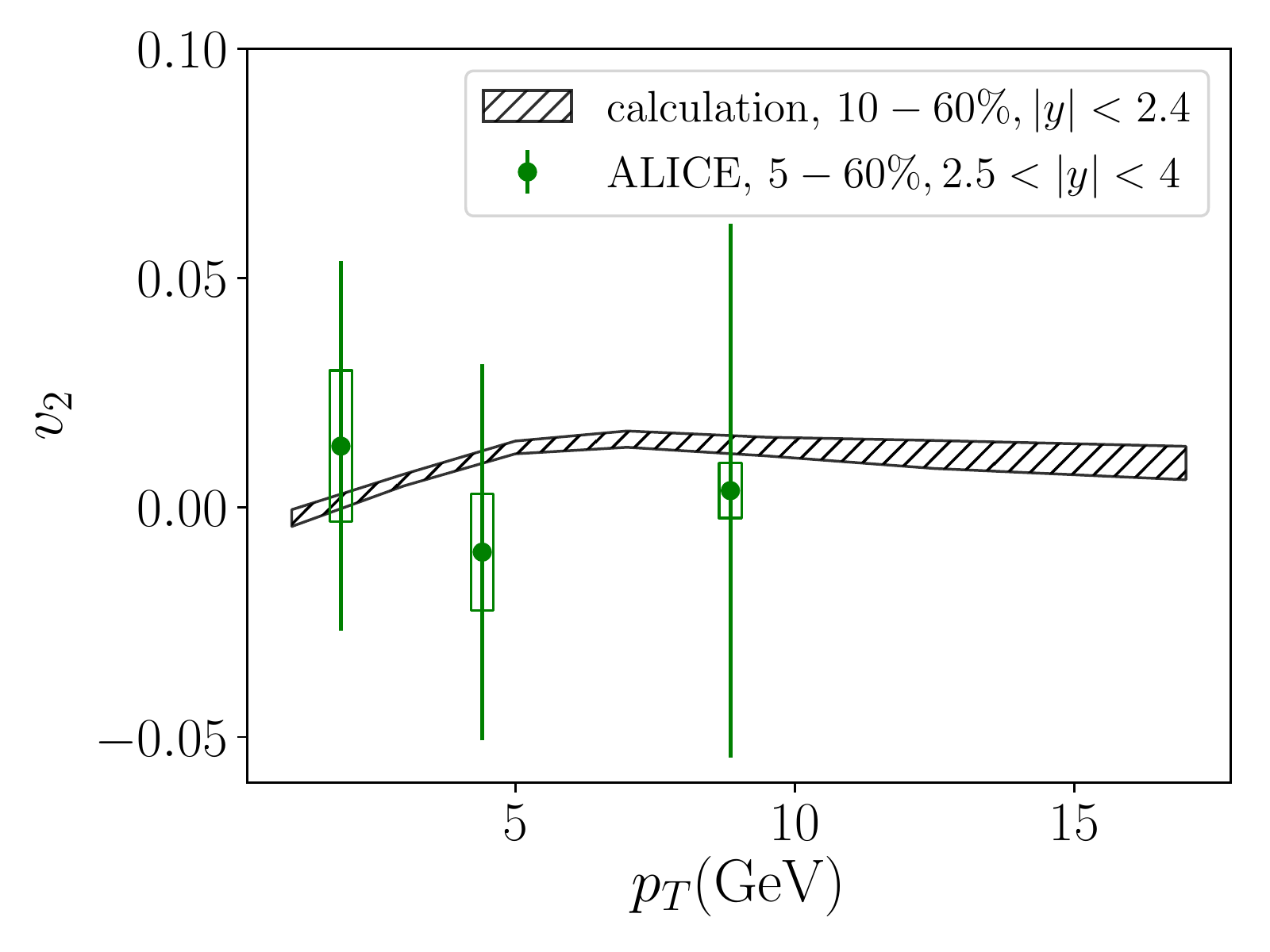}
    \caption{Results of $R_{AA}$ and $v_2$ in $5.02$ TeV Pb-Pb collisions compared with CMS measurements of $R_{AA}$ \cite{Sirunyan:2018nsz} and ALICE measurements of $v_2$ \cite{Acharya:2019hlv}. $N_{\rm{part}}$, $p_T$ and $y$ denote the collision centrality (related to the impact parameter), quarkonium transverse momentum and rapidity.}
    \label{fig:data_compare}
\end{figure}

In heavy ion experiments, two observables of quarkonium are of interest: the nuclear modification factor $R_{AA}$ and the azimuthal anisotropy coefficients $v_n$. The former is defined by the ratio of the quarkonium production cross section in heavy ion collisions and that in proton-proton collisions, scaled by the effective number of binary nucleon-nucleon collisions in heavy ion collisions such that $R_{AA}=1$ without any medium effect. The latter is defined by Fourier decomposition of the quarkonium azimuthal distribution into $\sum_n 2 v_n \cos(2n\phi)$, where $\phi$ is the azimuthal angle of quarkonium with respect to the reaction plane that is defined event-by-event. Our calculation results for $5.02$ TeV Pb-Pb collisions are shown in Fig.~\ref{fig:data_compare}, which are in good agreement with the experimental measurements.

\section{Conclusions}\label{aba:sec5}
We showed how to derive the quarkonium transport equation inside the QGP in a theoretically controlled way with a separation of scales. This provides a theoretical justification of applying transport equation in phenomenology of quarkonium production in heavy ion collisions. Phenomenological results on Upsilon production were also presented which agree well with data. Extension to the study of doubly heavy baryon can also be made \cite{Yao:2018zze}. Quarkonium transport equations in a different hierarchy of scales ($T\gg Mv^2$) have been discussed in Ref.~\citenum{Brambilla:2017zei,Miura:2019ssi}.

\section*{Acknowledgments}
This work is supported by U.S. Department of Energy research grants DE-FG02-05ER41367 and DE-FG02-05ER41368. XY also acknowledges support from DE-SC0011090, Brookhaven National Laboratory and Department of Physics, Massachusetts Institute of Technology.

\end{document}